# A Real-Valued Description of Quantum Mechanics with Schrödinger's 4th-order Matter-Wave Equation

**Nicos Makris[1] and Gary F. Dargush[2]**

[1]Civil and Environmental Engineering, Southern Methodist University, Dallas, Texas

[2]Mechanical and Aerospace Engineering, University at Buffalo, New York

**Abstract**

Using a variational formulation, we show that Schrödinger's 4th-order, real-valued matter-wave equation which involves the spatial derivatives of the potential $V(r)$, produces the precise eigenvalues of Schrödinger's 2nd-order, complex-valued matter-wave equation together with an equal number of negative, mirror eigenvalues. Accordingly, the paper concludes that there is a real-valued description of non-relativistic quantum mechanics in association with the existence of negative (repelling) energy levels. Schrödinger's classical 2nd-order, complex-valued matter-wave equation which was constructed upon factoring the 4th-order, real-valued differential operator and retaining only one of the two conjugate complex operators is a simpler description of the matter-wave, since it does not involve the derivatives of the potential $V(r)$, at the expense of missing the negative (repelling) energy levels.

**Introduction**

The question whether complex numbers are fundamental and indispensable in the mathematical description of quantum mechanics accompanied its development since its early stages. Upon E. Schrödinger developed his iconic 2nd-order in space and first-order in time, complex-valued matter-wave equation (equation (7) in this paper) which was published in [1]—also translated in English in [2], on June 06, 1926 he wrote a letter to H. A. Lorentz where he expressed his discomfort in his own words: "*What is unpleasant here, and indeed directly to be objected to, is the use of complex numbers. ψ is surely fundamentally a real function….*" [3]. Furthermore, at the end of his seminal part IV, 1926 paper [1], received by the publisher on June 23, 1926 [4], Schrödinger concludes: "*Meantime, there is no doubt a certain crudeness in the use of a complex wave function. If it were unavoidable in principle, and not merely a facilitation of the calculation, this would mean that there are actually two wave functions [ψ_r and ψ_i], which must be used together in order to obtain information on the state of the system. This somewhat unacceptable inference is to be replaced, I believe, of the very much more congenial interpretation that the state* *of the system is given by a real function and its time derivative.*" Schrödinger struggled for some time seeking a real-valued description of the matter-wave equation [4-8]. His initial goal was to remove the imaginary unit $i = \sqrt{-1}$, from the 2nd-order in space and first-order in time, complex-valued matter-wave equation (equation (7) in this paper); therefore, he introduced higher-order derivatives to reach a 4th-order in space and 2nd-order in time, real-valued matte-wave equation (equation (5) in this paper), which is of central interest in this present study.

While, in his 1926, part IV paper [1], Schrödinger explains that *"the 4th-order real-valued equation is thus evidently the uniform and general equation for the field scalar $\psi(\mathbf{r}, t)$"* it has received marginal attention [5, 7, 8] and its predictions were apparently never explored in depth. Within the course of a few months upon the publication of his 1926 paper [1], it appears that Schrödinger's discomfort with the 2nd-order in space and 1st--order in time, complex-valued matter-wave equation published in [1] was eventually removed and any attempt to develop a real-valued description by the pioneers of quantum mechanics was soon abandoned [4-9]. Nevertheless, the possibility of using a real-valued formulation for quantum mechanics was never formally ruled out. Chen [5, 7] showed that the real part of the wave function encodes the full information of the matter wave and derived a 4th-order real valued equation for a non-conservative system. A recent study [10] suggests that a real-valued quantum theory can describe a broad range of quantum systems; while, other recent studies that hinge upon symmetry conditions of real number pairs [11], the de Sitter algebra [12] or involve entangled qubits [13–15] conclude that quantum mechanics can only be described with complex-valued functions.

In this paper we implement a variational formulation and we show that Schrödinger's 4th-order, real-valued matter-wave equation which involves the spatial derivatives of the potential, $V(\mathbf{r})$ (equation (8) in this paper, which is an expanded expression of equation (5)), produces the precise eigenvalues of Schrödinger's 2nd-order, complex-valued matter-wave equation (which does not involve derivatives of the potential) together with an equal number of negative, mirror eigenvalues. First, we confirm that the known stationary states (Hermite polynomials) for the quantum harmonic oscillator, satisfy Schrödinger's 4th-order, real-valued matter-wave equation (8), which involves the derivatives of the potential $V(\mathbf{r})$; and subsequently we present a general variational formulation to compute the eigenvalues for any given potential, $V(\mathbf{r})$. Our variational formulation in association with a novel numerical scheme predicts the energy levels of the hydrogen atom with remarkable accuracy; while, it uncovers an equal number of negative, mirror eigenvalues. Accordingly, our paper concludes that there is a real-valued description of non-relativistic quantum mechanics in association with the co-existence of negative (repelling) energy levels.

**Schrödinger's original 4th-order, real-valued matter-wave equation**

In his part II, Annalen der Physik, Vol 79, 1926 paper [16], Schrödinger has reached the following time-independent, matter-wave equation (equation (18'') in [16]).

$$\frac{1}{m}\nabla^2\psi(r) + \frac{2}{\hbar^2}(E - V(r))\psi(r) = 0 \tag{1}$$

where $m$ is the mass of a non-relativistic elementary particle, $E$ is the energy- or frequency-parameter, $V(r)$, is the potential energy that is only a function of the position $r$ and $\hbar = h/2\pi$ where $h = 6.62607 \times 10^{-34}\ m^2kg/s =$ Plank's constant. Schrödinger explains in his part IV, 1926 paper [1], that the scalar $\psi(r)$ in equation (1) is merely a time independent amplitude (mode shape); therefore, equation (1) needs to be generalized in order to account for the time-dependence [4,6]. For a standing wave the spatial and temporal dependence of the matter-wave can be separated,

$$\psi(r,t) = \psi(r)e^{-\frac{i}{\hbar}Et} \tag{2}$$

so that

$$\frac{\partial \psi(r,t)}{\partial t} = -\frac{i}{\hbar}E\psi(r,t) \quad \text{and} \quad \frac{\partial^2 \psi(r,t)}{\partial t^2} = -\frac{E^2}{\hbar^2}\psi(r,t) \tag{3}$$

In his initial effort to remove the imaginary unit $i = \sqrt{-1}$, Schrödinger adopts the second of equation (3) and rewrites equation (1) in a quadratic form in which the differential and energy operators are acting once more from the left on the time-dependent scalar $\psi(r,t)$.

$$\left(\frac{1}{m}\nabla^2 - \frac{2}{\hbar^2}V(r)\right)\left(\frac{1}{m}\nabla^2 - \frac{2}{\hbar^2}V(r)\right)\psi(r,t) = \frac{4}{\hbar^2}\frac{E^2}{\hbar^2}\psi(r,t)$$

$$= -\frac{4}{\hbar^2}\left(-\frac{E^2}{\hbar^2}\psi(r,t)\right) \tag{4}$$

Substitution of the second of equation (3) into the last parenthesis of equation (4) gives

$$\left(\frac{1}{m}\nabla^2 - \frac{2}{\hbar^2}V(\boldsymbol{r})\right)^2 \psi(\boldsymbol{r},t) + \frac{4}{\hbar^2}\frac{\partial^2 \psi(\boldsymbol{r},t)}{\partial t^2} = 0 \qquad (5)$$

which is a 4th-order in space and a 2nd-order in time, real-valued matter-wave equation. In his part IV, 1926 paper [1], Schrödinger explains in his own words: *"equation (5) is thus evidently the uniform and general equation for the field scalar $\psi(\boldsymbol{r},t)$."* Equation (5) produces the required quadratic dispersion relation between the angular frequency $\omega$ and the wave number $\kappa$ ($\omega = \frac{\hbar}{2m}\kappa^2$) as dictated by Einstein's [17] quantized energy equation, $E = h\nu = \hbar\omega$ and de Broglie's [18] momentum-wave length relation, $p = h/\lambda = \hbar\kappa$. Schrödinger [1,2] further recognizes that his *real, uniform and general* matter-wave equation (5) resembles the 4th-order equations of motion that emerge from the theory of elasticity and references the governing equation of a vibrating plate. In the interest of simplifying the calculations in the eigenvalue analysis of equation (5) in association that $V(\boldsymbol{r})$ does not contain the time, Schrödinger [1,2] reverts to equation (4) which can be factorized as

$$\left(\frac{1}{m}\boldsymbol{\nabla}^2 - \frac{2}{\hbar^2}V(\boldsymbol{r}) + \frac{2}{\hbar^2}E\right)\left(\frac{1}{m}\nabla^2 - \frac{2}{\hbar^2}V(\boldsymbol{r}) - \frac{2}{\hbar^2}E\right)\psi(\boldsymbol{r},t) = 0 \qquad (6)$$

Inspired by the factorized form of his *"uniform and general"* 4th-order, real-valued equation (5) given by equation (6), in which the first parenthesis is essentially equation (1); in association with the structure of the Hamilton-Jacobi equation [19-24], Schrödinger reverts to the first of equation (3) which involves the imaginary unit $i = \sqrt{-1}$, and replaces $E\psi$ in equation (1) with $i\hbar\, \partial\psi(r,t)/\partial t$. In this way, equation (1) becomes a time-dependent equation,

$$i\hbar\, \frac{\partial \psi(\boldsymbol{r},t)}{\partial t} = -\frac{\hbar^2}{2m}\nabla^2 \psi(\boldsymbol{r},t) + V(\boldsymbol{r})\psi(r,t) \qquad (7)$$

which is Schrödinger's iconic 2nd-order in space and 1st-order in time complex-valued wave equation [1,2], which predicts with remarkable accuracy the energy levels observed in the atomic line-spectra of the chemical elements in association with other features of the periodic table [25-29].

While Schrodinger's 2nd-order, complex-valued equation (9) does not contain spatial derivatives of the potential, $V(\boldsymbol{r})$, his 4th-order, real-valued equation (5) involves spatial derivatives of the potential $V(\mathbf{r})$, as indicated with the succession of differential operators appearing in equation (4), which in its expanded form assumes the time-dependent expression

$$-\hbar^2 \frac{\partial^2 \psi(\mathbf{r},t)}{\partial t^2} = \frac{\hbar^4}{4m^2}\nabla^4\psi(\mathbf{r},t) - \frac{\hbar^2}{m}V(\mathbf{r})\nabla^2\psi(\mathbf{r},t) - \frac{\hbar^2}{m}\nabla V(\mathbf{r})\nabla\psi(\mathbf{r},t) + \left(V^2(\mathbf{r}) - \frac{\hbar^2}{2m}\nabla^2 V(\mathbf{r})\right)\psi(\mathbf{r},t) \quad (8)$$

Returning to equation (4), the time-independent 4$^{th}$-order Schrödinger equation is

$$\frac{\hbar^4}{4m}\nabla^4\psi(\mathbf{r}) - \frac{\hbar^2}{m}V(\mathbf{r})\nabla^2\psi(\mathbf{r}) - \frac{\hbar^2}{m}\nabla V(\mathbf{r})\nabla\psi(\mathbf{r}) + \left(V^2(\mathbf{r}) - \frac{\hbar^2}{2m}\nabla^2 V(\mathbf{r})\right)\psi(\mathbf{r}) = E^2\psi(\mathbf{r}) \quad (9)$$

In a recent publication, Makris [30] explored the predictions of equation (5) when an elementary non-relativistic particle is trapped in a finite potential square well. When expanding the quadratic term of equation (5), he noticed that if the differential operator, $\nabla^2$, in the cross-term, $\frac{4}{m}\nabla^2 V(\mathbf{r})$, is on the right of V(r) (that is to avoid to take spatial derivatives of the potential), then equation (5) bears striking similarities with the equation of a flexural-shear beam supported on an elastic foundation. Inspired from this analogy, Makris [30] presented an analysis for the energy levels of a particle trapped in a finite potential, square well using Schrödinger's 4$^{th}$-order equation (5) without taking the spatial derivatives of the potential (Dirac delta functions and its derivatives at the walls of the well) which was incorrect. The correct expression of the expanded form of Schrödinger's 4$^{th}$-order real-valued equation (5) is equation (8) which involves spatial derivatives of the potential V(r).

**Energy levels of a particle trapped in a quantum harmonic potential**

For an elementary particle with mass, $m$, trapped in a one-dimensional, quantum harmonic well, the potential $V(x) = kx^2/2 = m\omega^2 x^2/2$, where $\omega = \sqrt{k/m}$, is the frequency of the harmonic oscillator. Then, the one-dimensional version of the time-independent 2$^{nd}$- and 4$^{th}$-order matter-wave equations derive from equations (1) and (9), respectively, producing

$$-\frac{\hbar^2}{2m}\frac{d^2\psi(x)}{dx^2} + \frac{1}{2}m\omega^2 x^2 \psi(x) = E\psi(x) \tag{10}$$

and

$$\frac{\hbar^4}{4m^2}\frac{d^4\psi(x)}{dx^4} - \frac{\hbar^2\omega^2}{2}x^2\frac{d^2\psi(x)}{dx^2} - \hbar^2\omega^2 x\frac{d\psi(x)}{dx} + \left(\frac{1}{4}m^2\omega^4 x^4 - \frac{\hbar^2\omega^2}{2}\right)\psi(x) = E^2\psi(x) \tag{11}$$

The reader can verify that the stationary states for the quantum harmonic oscillator which are derived upon integrating Schrodinger's 2$^{nd}$-order, time-independent equation [31-33]

$$\psi_n(x) = C_n H_n\left(\sqrt{\frac{m\omega}{\hbar}}x\right)e^{-\frac{m\omega}{2\hbar}x^2}, \quad n \in \{0,1,2,3\ldots\}, \tag{12}$$

where $H_n(z)$ are the Hermite polynomials ($H_0 = 1$, $H_1 = 2z$, $H_2 = 4z^2 - 2$, ….) and $C_n$ is a normalization constant. For orthonormality over the entire $x$-axis, $C_0 = 1/(\sqrt[4]{\pi})$, $C_1 = 1/(\sqrt{2}\sqrt[4]{\pi})$ and $C_2 = 1/(2\sqrt{2}\sqrt[4]{\pi})$ for the first three modes. These are also solutions of the 4$^{th}$-order equation (11), which involves the derivatives of the potential $V(x) = m\omega^2 x^2/2$ and yields twice as many energy levels:

$$\left(n+\frac{1}{2}\right)^2 \hbar^2\omega^2 = E_n^2 \Rightarrow E_n = \pm\left(n+\frac{1}{2}\right)\hbar\omega \tag{13}$$

Accordingly, for any $n$, equation (13) yields the known energy levels, $E_{n+} = \left(n+\frac{1}{2}\right)\hbar\omega$ that result from Schrodinger's 2$^{nd}$-order, complex-valued equation (7), together with a mirror, negative energy level, $E_{n-} = -\left(n+\frac{1}{2}\right)\hbar\omega$. For instance, for $n = 0$, substitution of the corresponding eigenfunction, $\psi_0(x) = C_0 e^{-\frac{m\omega}{2\hbar}x^2}$, given by equation (12) into equation (11) yields $\frac{1}{4}\hbar^2\omega^2 = E_0^2$; therefore, $E_0 = \pm\frac{1}{2}\hbar\omega$. Similarly, for $n = 1$, substitution of the corresponding eigenfunction, $\psi_1(x) = 2C_1\sqrt{\frac{m\omega}{\hbar}}xe^{-\frac{m\omega}{2\hbar}x^2}$, given by equation (12) into equation (11) yields $\frac{9}{4}\hbar^2\omega^2 = E_1^2$ ; therefore, $E_1 = \pm\frac{3}{2}\hbar\omega$ and so forth.

**Variational formulations for the quantum harmonic potential**

An attractive approach to find solutions to eigenproblems in mathematical physics is to develop a variational formulation. Ramdas Ram-Mohan [34] presented such formulations for the quantum harmonic potential and the radial modes of the hydrogen atom, based on the Schrodinger 2nd order equation. Subsequently, he developed a corresponding finite element method using linear elemental shape functions to obtain approximate solutions for the lowest few modes. We begin in this section with a similar formulation for the 2nd-order Schrodinger equation for the harmonic potential and then extend the approach to consider the 4th-order, real-valued matter-wave equation (11).

The second order time-independent Schrodinger equation (10) for the harmonic potential can be rewritten in terms of a non-dimensional coordinate $\xi$ as follows:

$$\frac{d^2\psi(\xi)}{d\xi^2} - \xi^2\psi(\xi) + \epsilon\psi(\xi) = 0 \tag{14}$$

with dimensionless eigenenergy $\epsilon = 2E/\hbar\omega$. This represents the strong (or differential) form of the problem.

A variational statement can be developed by multiplying equation (14) by a virtual wave function $\delta\psi^*(\xi)$, where the symbol $*$ indicates complex conjugate, and then integrating over the entire $\xi$-axis. As a result, the following virtual correlated action can be written:

$$\delta\mathcal{A}_{2h} = \int_{-\infty}^{\infty} \delta\psi^*(\xi) \left\{ \frac{d^2\psi(\xi)}{d\xi^2} - \xi^2\psi(\xi) + \epsilon\psi(\xi) \right\} d\xi \tag{15}$$

Applying integration-by-parts on the first term in (15) leads to the following for the stationary correlated action:

$$\delta\mathcal{A}_{2h} = \int_{-\infty}^{\infty} \left\{ -\frac{d\delta\psi^*(\xi)}{d\xi}\frac{d\psi(\xi)}{d\xi} - \delta\psi^*(\xi)\xi^2\psi(\xi) + \epsilon\,\delta\psi^*(\xi)\psi(\xi) \right\} d\xi$$
$$+ \delta\psi^*(\xi)\frac{d\psi(\xi)}{d\xi}\Big|_{-\infty}^{\infty} = 0 \tag{16}$$

Assuming the derivative of the wave function decays as $\xi \to \pm\infty$, the boundary terms vanish and after reversing the signs, the weak (or integral) form of the 2$^{nd}$ order harmonic operator reduces to

$$\delta\mathcal{A}_{2h} = \int_{-\infty}^{\infty} \left\{ \frac{d\delta\psi^*(\xi)}{d\xi}\frac{d\psi(\xi)}{d\xi} + \delta\psi^*(\xi)\xi^2\psi(\xi) - \epsilon\,\delta\psi^*(\xi)\psi(\xi) \right\} d\xi = 0 \tag{17}$$

Equation (17) involves at most first order derivatives and thus represents a $C^0$-variational problem, which will only require continuity of the wave function $\psi(\xi)$ within a computational method. Notice that each term is symmetric with respect to $\delta\psi^*(\xi)$ and $\psi(\xi)$.

Using a similar procedure, a variational statement can be developed for the fourth order time-independent Schrodinger equation (11) for the harmonic potential. Introducing the non-dimensional coordinate $\xi$ and eigenenergy $\epsilon$ as defined above, equation (11) can be rewritten in the following non-dimensional strong form:

$$\frac{d^4\psi(\xi)}{d\xi^4} - 2\xi^2\frac{d^2\psi(\xi)}{d\xi^2} - 4\xi\frac{d\psi(\xi)}{d\xi} + (\xi^4 - 2)\psi(\xi) + \epsilon^2\,\psi(\xi) = 0 \tag{18}$$

In this case, the corresponding virtual correlated action for the 4$^{th}$-order harmonic operator becomes

$$\delta\mathcal{A}_{4h} = \int_{-\infty}^{\infty} \delta\psi^*(\xi) \left\{ \frac{d^4\psi(\xi)}{d\xi^4} - 2\xi^2\frac{d^2\psi(\xi)}{d\xi^2} - 4\xi\frac{d\psi(\xi)}{d\xi} + (\xi^4 - 2)\psi(\xi) \right.$$
$$\left. + \epsilon^2\,\psi(\xi) \right\} d\xi \tag{19}$$

Again, integration-by-parts is needed to reduce the order of derivatives in equation (19) to create the weak form. However, for the 4$^{th}$-order Schrödinger wave equation, integration-by-parts must be applied twice on the first term. The final form, after enforcing the decay of boundary terms as $\xi \to \pm\infty$, can be written

$$\delta \mathcal{A}_{4h} = \int_{-\infty}^{\infty} \left\{ \frac{d^2 \delta\psi^*(\xi)}{d\xi^2} \frac{d^2 \psi(\xi)}{d\xi^2} + \frac{d\delta\psi^*(\xi)}{d\xi} 2\xi^2 \frac{d\psi(\xi)}{d\xi} + \delta\psi^*(\xi)(\xi^4 - 2)\psi(\xi) \right. \\ \left. - \epsilon^2 \, \delta\psi^*(\xi) \, \psi(\xi) \right\} d\xi = 0 \qquad (20)$$

Once again, each term in equation (20) displays symmetry with respect to $\delta\psi^*(\xi)$ and $\psi(\xi)$. Now, for the 4$^{th}$-order Schrodinger equation, the weak form contains second order derivative and thus becomes a $C^1$-variational problem that requires continuity of the wave function $\psi(\xi)$ and its first derivative $d\psi(\xi)/d\xi$ for the convergence of numerical solutions based on equation (20).

Alternatively, a variational approach can be developed by seeking the stationarity of a correlated action $\tilde{\mathcal{A}}$, where the 2$^{nd}$- and 4$^{th}$-order versions can be written, respectively, as

$$\tilde{\mathcal{A}}_{2h} = \int_{-\infty}^{\infty} \frac{1}{2} \left\{ -\frac{d\psi^*(\xi)}{d\xi} \frac{d\psi(\xi)}{d\xi} - \psi^*(\xi)\xi^2 \psi(\xi) + \epsilon \, \psi^*(\xi)\psi(\xi) \right\} d\xi \qquad (21)$$

and

$$\tilde{\mathcal{A}}_{4h} = \int_{-\infty}^{\infty} \left\{ \frac{d^2 \psi^*(\xi)}{d\xi^2} \frac{d^2 \psi(\xi)}{d\xi^2} + \frac{d\psi^*(\xi)}{d\xi} 2\xi^2 \frac{d\psi(\xi)}{d\xi} + \psi^*(\xi)(\xi^4 - 2)\psi(\xi) \right. \\ \left. - \epsilon^2 \, \psi^*(\xi) \, \psi(\xi) \right\} d\xi \qquad (22)$$

by ignoring all boundary terms. Taking first variations of equations (21) and (22) produces the following weak forms:

$$\delta \tilde{\mathcal{A}}_{2h} = \int_{-\infty}^{\infty} \left\{ -\delta\left(\frac{d\psi^*(\xi)}{d\xi}\right) \frac{d\psi(\xi)}{d\xi} - \delta\psi^*(\xi)\xi^2 \psi(\xi) + \epsilon \, \delta\psi^*(\xi)\psi(\xi) \right\} d\xi = 0 \qquad (23)$$

and

$$\delta \tilde{\mathcal{A}}_{4h} = \int_{-\infty}^{\infty} \left\{ \delta\left(\frac{d^2\psi^*(\xi)}{d\xi^2}\right)\frac{d^2\psi(\xi)}{d\xi^2} + \delta\left(\frac{d\psi^*(\xi)}{d\xi}\right)2\xi^2\frac{d\psi(\xi)}{d\xi} + \delta\psi^*(\xi)(\xi^4 - 2)\psi(\xi) \right. \\ \left. - \epsilon^2\,\delta\psi^*(\xi)\,\psi(\xi) \right\} d\xi = 0 \qquad (24)$$

which are identical to (17) and (20), except for the reversal in order of variations and derivatives. Integration-by-parts operations on these latter two equations will recover (14) and (18), respectively, as the Euler-Lagrange equations of the correlated actions $\tilde{\mathcal{A}}_{2h}$ and $\tilde{\mathcal{A}}_{4h}$.

**Variational formulations for radial modes of the hydrogen atom**

For the hydrogen atom, the potential $V(r) = e^2/r = \hbar^2/mr$ [31-33] and the second order, time-independent Schrodinger equation (1) for radial modes can be written in terms of a non-dimensional radial coordinate $\rho$ as follows:

$$\nabla^2 \psi(\rho) + \frac{2}{\rho}\psi(\rho) + \epsilon\psi(\rho) = 0 \qquad (25)$$

where $\nabla^2$ *is* the Laplace operator in radial coordinates, such that

$$\nabla^2 = \frac{1}{\rho^2}\frac{d}{d\rho}\left(\rho^2 \frac{d}{d\rho}\right) \qquad (26)$$

while $\epsilon = E/R_\infty$ with $R_\infty = 13.6\text{eV}$ [31-33]. Then, the virtual correlated action can be written

$$\delta \hat{\mathcal{A}}_{2H} = \int_0^\infty \delta\psi^*(\rho) \left\{ \frac{1}{\rho^2}\frac{d}{d\rho}\left(\rho^2 \frac{d\psi(\rho)}{d\rho}\right) + \frac{2}{\rho}\psi(\rho) + \epsilon\psi(\rho) \right\} 4\pi\rho^2 d\rho \qquad (27)$$

For stationarity, and after cancelling the $4\pi$ factor, this becomes

$$\delta\mathcal{A}_{2H} = \int_0^\infty \delta\psi^*(\rho)\left\{\frac{d}{d\rho}\left(\rho^2\frac{d\psi(\rho)}{d\rho}\right) + 2\rho\psi(\rho) + \epsilon\,\rho^2\psi(\rho)\right\}d\rho = 0 \qquad (28)$$

By performing integration-by-parts on the first term in (28), the following weak form can be established

$$\int_0^\infty \left\{\frac{d\delta\psi^*(\rho)}{d\rho}\rho^2\frac{d\psi(\rho)}{d\rho} - \delta\psi^*(\rho)\,2\rho\,\psi(\rho) - \epsilon\,\delta\psi^*(\rho)\rho^2\psi(\rho)\right\}d\rho = 0 \qquad (29)$$

after reversing the signs and assuming that the boundary terms vanish. As was the case for the 2$^{nd}$ order harmonic oscillator formulation, all terms in (29) are symmetric with respect to $\delta\psi^*(\rho)$ and $\psi(\rho)$ and involve at most first order derivatives. Consequently, this also is a $C^0$-variational problem. Note that Eq. (29) is equivalent to the weak form that would result from a variation of the action integral given by Ramdas Ram-Mohan [34].

In non-dimensional form, the 4$^{th}$-order Schrodinger equation for the radial response for the hydrogen atom can be written as

$$\left(\nabla^2 + \frac{2}{\rho} - \epsilon\right)\left(\nabla^2 + \frac{2}{\rho} + \epsilon\right)\psi(\rho) = 0 \qquad (30)$$

As a result, the corresponding virtual correlated action becomes

$$\delta\hat{\mathcal{A}}_{4H} = \int_0^\infty \delta\psi^*(\rho)\left\{\left(\nabla^2 + \frac{2}{\rho} - \epsilon\right)\left(\nabla^2 + \frac{2}{\rho} + \epsilon\right)\psi(\rho)\right\}4\pi\rho^2 d\rho \qquad (31)$$

or after expanding the composition of the second order operators, this converts to the following form under the condition for stationarity:

$$\delta\hat{\mathcal{A}}_{4H} = \int_0^\infty \delta\psi^*(\rho)\left\{\nabla^4\psi(\rho) + \nabla^2\left(\frac{2}{\rho}\psi(\rho)\right) + \frac{2}{\rho}\nabla^2\psi(\rho) + \frac{4}{\rho^2}\psi(\rho)\right. \tag{32}$$
$$\left. - \epsilon^2\,\psi(\rho)\right\}\rho^2 d\rho = 0$$

However,

$$\nabla^2\left(\frac{2}{\rho}\psi(\rho)\right) = \nabla^2\left(\frac{2}{\rho}\right)\psi(\rho) + 2\nabla\left(\frac{2}{\rho}\right)\nabla\psi(\rho) + \frac{2}{\rho}\nabla^2\psi(\rho) \tag{33}$$

$$\int_\Omega \nabla^2\left(\frac{1}{\rho}\right)f(\rho)d\Omega = \int_0^\infty \nabla^2\left(\frac{1}{\rho}\right)f(\rho)4\pi\rho^2 d\rho = -4\pi\int_0^\infty \delta(\rho)f(\rho)d\rho \tag{34}$$

$$2\nabla\left(\frac{2}{\rho}\right)\nabla\psi(\rho) = -\frac{4}{\rho^2}\frac{d\psi(\rho)}{d\rho} \tag{35}$$

where $\delta(\rho)$ is the Dirac delta function. After substituting equations (33)-(35) and cancelling out the common factor $4\pi$, equation (32) becomes

$$\delta\mathcal{A}_{4H} = \int_0^\infty \delta\psi^*(\rho)\left\{\nabla^4\psi(\rho) + \frac{4}{\rho}\nabla^2\psi(\rho) - \frac{4}{\rho^2}\frac{d\psi(\rho)}{d\rho} + \frac{4}{\rho^2}\psi(\rho) - \frac{2}{\rho^2}\delta(\rho)\psi(\rho)\right. \tag{36}$$
$$\left. - \epsilon^2\,\psi(\rho)\right\}\rho^2 d\rho = 0$$

Applying integration-by-parts to the first two terms in equation (36) provides the following weak form for the 4$^{th}$ order real-valued matter-wave equation:

$$\int_0^\infty \left\{ \frac{d^2 \delta\psi^*(\rho)}{d\rho^2} \rho^2 \frac{d^2\psi(\rho)}{d\rho^2} + \frac{d^2 \delta\psi^*(\rho)}{d\rho^2} 2\rho \frac{d\psi(\rho)}{d\rho} + \frac{d\delta\psi^*(\rho)}{d\rho} 2\rho \frac{d^2\psi(\rho)}{d\rho^2} \right. $$
$$\left. + \frac{d\delta\psi^*(\rho)}{d\rho} 4(1-\rho) \frac{d\psi(\rho)}{d\rho} + \delta\psi^*(\rho)\, 4\, \psi(\rho) - \epsilon\, \delta\psi^* \rho^2\, \psi(\rho) \right\} d\rho \qquad (37)$$
$$- \delta\psi^*(0)\, 2\, \psi(0) = 0$$

Notice once again the symmetry of the formulation with respect to $\delta\psi^*(\rho)$ and $\psi(\rho)$. Also, second order derivatives are present in the weak form, thus requiring a $C^1$ representation involving continuity of the wave function, along with its first derivative.

**Numerical solutions of the variational forms based on a Ritz Spline Method (RSM)**

The four variational statements provided in the previous sections, given by Eqs. (17), (20), (29) and (37), are symmetric weak forms with respect to the virtual $\delta\psi^*(\rho)$ and actual $\psi(\rho)$ wave functions. Consequently, discrete representations of these forms can be written exclusively in terms of real, symmetric system matrices and lead to matrix eigenproblems with real eigenvalues and sets of corresponding real orthogonal eigenvectors. Complex algebra is not necessary and the complex virtual wave functions $\delta\psi^*(\rho)$ can be replaced by their real counterparts $\delta\psi(\rho)$.

Both weak forms, Eqs. (17) and (29), for the 2nd order Schrodinger equations require $C^0$ continuity, meaning that only the wave functions need to be represented as continuous in the numerical approximation. Meanwhile, $C^1$ continuity must be enforced for the 4th-order weak formulations in Eqs. (20) and (37). For these cases, both the wave functions and their first spatial derivatives must be continuous.

There are several alternative approaches that could be used to solve the eigenproblems associated with the four weak forms. For example, standard finite element methods are applicable, as demonstrated by Ramdas Ram-Mohan [34] for the 2nd-order Schrodinger equations. However, a Ritz Spline Method (RSM), introduced in [35, 36] for size-dependent couple stress problems of continuum mechanics, is preferable due to the flexibility in the selection of the spline function order and level of continuity. Additionally, an extension to multi-dimensional problems is straightforward [35, 36].

Here, more specifically, we let the wave function be represented in terms of real, univariate B-spline basis functions of order $k - 1$ having $C^m$ continuity, where $k$ is the number of coefficients in each spline. Thus, we take

$$\psi(\xi) \cong \langle N(\xi)|c\rangle \tag{38}$$

where $\xi$ is the local intrinsic coordinate, the row vector $\langle N(\xi)|$ consists of the polynomial spline functions over a finite range of $\xi$, and $|c\rangle$ is the column vector of $n$ unknown spline coefficients. For the harmonic potential, $\xi \in [-X, X]$, while $\xi \in [0, R]$ for the hydrogen atom.

Then, the first derivatives of the wave function can be approximated, as follows:

$$\frac{d\psi(\xi)}{d\xi} \cong \langle B(\xi)|c\rangle \tag{39}$$

where

$$\langle B(\xi)| = \frac{d\langle N(\xi)|}{d\xi} \tag{40}$$

for spline representations having $m \geq 0$. When $m \geq 1$, the second derivatives also can be evaluated via the relation

$$\frac{d^2\psi(\xi)}{d\xi^2} \cong \langle H(\xi)|c\rangle \tag{41}$$

with

$$\langle H(\xi)| = \frac{d\langle B(\xi)|}{d\xi} \tag{42}$$

To maintain symmetry of the discrete matrix variational representation, the virtual wave function must be defined in terms of identical splines. Thus, we let

$$\delta\psi(\xi) \cong \langle N(\xi)|\delta c\rangle \tag{43}$$

$$\frac{d\delta\psi(\xi)}{d\xi} \cong \langle B(\xi)|\delta c\rangle \tag{44}$$

and where applicable with $m \geq 1$,

$$\frac{d^2\delta\psi(\xi)}{d\xi^2} \cong \langle H(\xi)|\delta c\rangle \tag{45}$$

for $|\delta c\rangle$ as the vector of $n$ virtual spline coefficients.

Next, the discretized versions of the individual weak forms are obtained by substituting Eqs. (38)-(45) and performing the required integrations. Note that all four RSM formulations to be presented in the next two sections reduce to solutions of generalized eigenproblems involving real, sparse, symmetric matrices, to which very efficient computational algorithms are applied.

**Ritz Spline Method (RSM) for the quantum harmonic potential**

We first consider the second order time-independent Schrodinger equation for the quantum harmonic potential, represented by the weak form presented in Eq. (17), requiring only $C^0$ continuity. After substituting the spline representations from Eqs. (38), (39), (43) and (44) into the weak form

$$\delta \mathcal{A}_{2h} \cong \int_{-X}^{X} \{-\langle \delta c|B(\xi)\rangle\langle B(\xi)|c\rangle - \xi^2\langle \delta c|N(\xi)\rangle\langle N(\xi)|c\rangle + \epsilon\langle \delta c|N(\xi)\rangle\langle N(\xi)|c\rangle\} d\xi = 0 \tag{46}$$

After recognizing that the spline coefficients are independent of the intrinsic coordinate, Eq. (46) can be written

$$\langle \delta c| \int_{-X}^{X} \{|B(\xi)\rangle\langle B(\xi)| + \xi^2|N(\xi)\rangle\langle N(\xi)| - \epsilon|N(\xi)\rangle\langle N(\xi)|\} d\xi \, |c\rangle = 0 \tag{47}$$

or

$$\langle \delta c| \, [\mathbb{K}_{2h} - \epsilon\, \mathbb{M}_{2h}] \, |c\rangle = 0 \tag{48}$$

with real symmetric matrices

$$\mathbb{K}_{2h} = \int_{-X}^{X} \{|B(\xi)\rangle\langle B(\xi)| + \xi^2 |N(\xi)\rangle\langle N(\xi)|\} d\xi \tag{49}$$

$$\mathbb{M}_{2h} = \int_{-X}^{X} |N(\xi)\rangle\langle N(\xi)| d\xi \tag{50}$$

For arbitrary virtual spline coefficients $|\delta c\rangle$, the scalar discretized weak form Eq. (48) leads to the following matrix eigenproblem:

$$[\mathbb{K}_{2h} - \epsilon\, \mathbb{M}_{2h}] |c\rangle = |0\rangle \tag{51}$$

for real eigenvalues $\epsilon$ and an orthogonal set of eigenvectors $|c\rangle$, written in terms of real spline coefficients. Note that both $\mathbb{K}_{2h}$ and $\mathbb{M}_{2h}$ are symmetric, positive definite. Consequently, all eigenvalues for the 2nd order harmonic oscillator problem are positive.

The fourth order weak form for the harmonic potential requires $C^1$ continuity due to the presence of second order derivatives in eq. (20). Consequently, all the discretized relations Eqs. (38)-(45) are needed to form the approximation for $\delta \mathcal{A}_{4h}$. After making those substitutions,

$$\delta \mathcal{A}_{4h} \cong \int_{-X}^{X} \{\langle \delta c|H(\xi)\rangle\langle H(\xi)|c\rangle + 2\xi^2 \langle \delta c|B(\xi)\rangle\langle B(\xi)|c\rangle \\ + (\xi^4 - 2)\langle \delta c|N(\xi)\rangle\langle N(\xi)|c\rangle - \epsilon^2 \langle \delta c|N(\xi)\rangle\langle N(\xi)|c\rangle\} d\xi = 0 \tag{52}$$

or more succinctly after bringing the spline coefficients outside the integral

$$\langle \delta c| \int_{-X}^{X} \{|H(\xi)\rangle\langle H(\xi)| + 2\xi^2 |B(\xi)\rangle\langle B(\xi)| + (\xi^4 - 2)|N(\xi)\rangle\langle N(\xi)| \\ - \epsilon^2 |N(\xi)\rangle\langle N(\xi)|\} d\xi\, |c\rangle = 0 \tag{53}$$

Again, this can be written in simplified notation, as

$$\langle \delta c| [\mathbb{K}_{4h} - \epsilon^2\, \mathbb{M}_{4h}] |c\rangle = 0 \tag{54}$$

in terms of the real, symmetric matrices

$$\mathbb{K}_{4h} = \int_{-X}^{X} \{|H(\xi)\rangle\langle H(\xi)| + 2\xi^2|B(\xi)\rangle\langle B(\xi)| + (\xi^4 - 2)|N(\xi)\rangle\langle N(\xi)|\} d\xi \qquad (55)$$

$$\mathbb{M}_{4h} = \int_{-X}^{X} |N(\xi)\rangle\langle N(\xi)| d\xi \qquad (56)$$

With arbitrary virtual spline coefficients $|\delta c\rangle$, this leads to the following generalized matrix eigenproblem:

$$[\mathbb{K}_{4h} - \epsilon^2 \mathbb{M}_{4h}] |c\rangle = |0\rangle \qquad (57)$$

for real eigenvalues $\epsilon^2$ with an orthogonal set of eigenvectors $|c\rangle$. Note that $\mathbb{K}_{4h}$ may not be positive definite due to the final term Eq. (55). Consequently, all eigenvalues $\epsilon^2$ for the 4$^{th}$ order harmonic oscillator problem cannot be guaranteed to be positive.

For the numerical solution of the 2$^{nd}$-order Schrodinger equation for the quantum harmonic oscillator using the Ritz Spline Method (RSM) described in Eq. (51), we let the size of the computational domain stretch from $[-X, X]$ with $X = 10$. A total of $N = 100$ intervals of quartic splines ($k = 5$) having $C^1$ continuity are used. Nearly exact values of the non-dimensional eigenenergies are obtained, as shown in Table 1. Furthermore, in the 2$^{nd}$-order formulation the matrices are both analytically and numerically positive definite.

The 4$^{th}$-order Schrodinger matter-wave equation is solved based on the RSM presented in Eq. (57). Again, $X = 10$ and quartic splines ($k = 5$) with $C^1$ continuity are selected. However, convergence is more difficult with this higher order governing equation, so that a total of $N = 200$ spline intervals are applied to obtain the eigenenergies presented in the last column of Table 1. Notice once again that the ten lowest eigenenergies are calculated to at least eight significant digits. Furthermore, although the variational form is not necessarily positive definite, in the numerical simulations, all eigenvalues $\epsilon^2$ are positive, resulting the energy levels, $\epsilon$, computed with the 2$^{nd}$-order Schrodinger matter-wave equation together with an equal number of mirror, negative energy levels.

Figure 1 illustrates the first three eigenmodes for the RSM solutions in comparison to the analytical solutions to the second order Schrodinger equation given by equation (12). The modes are scaled to provide an orthonormal set in all three cases: the 2$^{nd}$-order Schrodinger analytical solutions, the 2$^{nd}$-order Schrodinger RSM solutions, and the 4$^{th}$-order Schrodinger RSM solutions. The well-known orthonormalized analytical eigenmodes [31-33] are defined in Eq. (12). For the RSM cases, the modes in orthonormal form satisfy

$$\langle c_i | \: \mathbb{M}_{nh} \: | c_j \rangle = \delta_{ij} \tag{58}$$

to many significant digits, where $|ci\rangle$ represents the spline coefficient column vector for the $i^{th}$ mode, $n$ denotes either the second or fourth order Schrodinger formulation and $\delta_{ij}$ is the Kronecker delta.

All these RSM results suggest that the variational form for the 2$^{nd}$ and 4$^{th}$-order Schrodinger equations produce nearly identical eigensolutions.

Table 1. Eigenenergies of a particle trapped in a quantum harmonic potential computed with Schrödinger's 2$^{nd}$-order and 4$^{th}$-order matter-wave equations

| Mode Number | Eigenenergy $\epsilon$ (2$^{nd}$ Order Exact) | Eigenenergy $\epsilon$ (2$^{nd}$ Order RSM) | Eigenvalue $\epsilon^2$ (4$^{th}$ Order RSM) | Eigenenergy $\epsilon$ (4$^{th}$ Order RSM) |
|---|---|---|---|---|
| 1 | 1 | 1.00000000 | 1.00000000 | $\pm$1.00000000 |
| 2 | 3 | 3.00000000 | 9.00000000 | $\pm$3.00000000 |
| 3 | 5 | 5.00000000 | 25.00000000 | $\pm$5.00000000 |
| 4 | 7 | 7.00000000 | 49.00000000 | $\pm$7.00000000 |
| 5 | 9 | 9.00000000 | 81.00000018 | $\pm$9.00000001 |
| 6 | 11 | 11.00000001 | 121.0000004 | $\pm$11.00000002 |
| 7 | 13 | 13.00000001 | 169.0000010 | $\pm$13.00000004 |
| 8 | 15 | 15.00000002 | 225.0000021 | $\pm$15.00000007 |
| 9 | 17 | 17.00000004 | 289.0000037 | $\pm$17.00000011 |
| 10 | 19 | 19.00000007 | 361.0000065 | $\pm$19.00000017 |

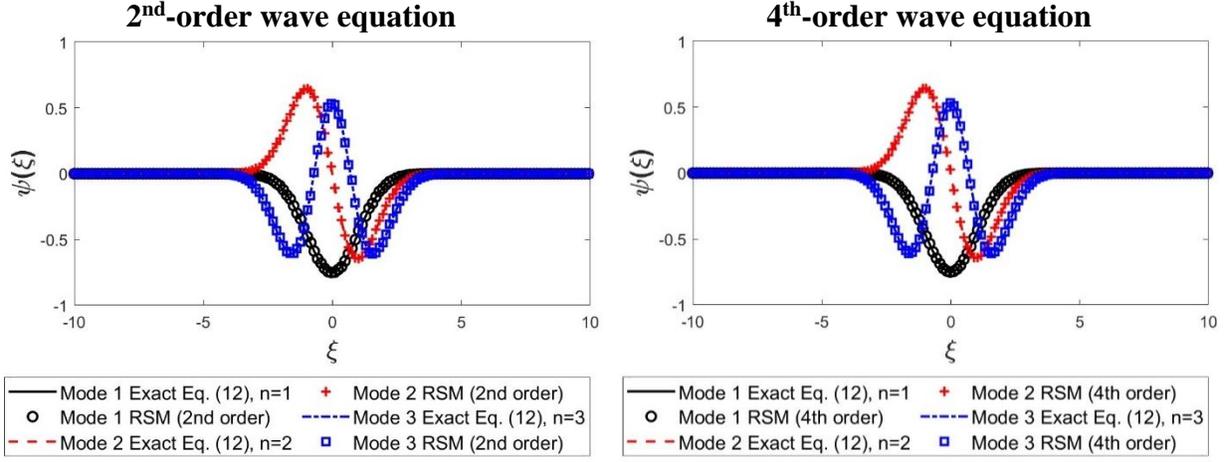

Figure 1. Eigenmodes of a particle trapped in a quantum harmonic potential. Exact 2nd-order solutions (lines) versus the computational results with the Ritz Spline Method (RSM-points) implemented to the 2nd order and 4th order descriptions

**Ritz Spline Method (RSM) for the hydrogen atom**

Starting with the weak form for the second order time-independent Schrodinger equation in Eq. (29), we can write the discretized version as follows:

$$\delta \mathcal{A}_{2H} \cong \int_0^R \{\rho^2 \langle \delta c|B(\rho)\rangle\langle B(\rho)|c\rangle - 2\rho \langle \delta c|N(\rho)\rangle\langle N(\rho)|c\rangle \\ - \epsilon\, \rho^2 \langle \delta c|N(\rho)\rangle\langle N(\rho)|c\rangle \} d\rho = 0 \tag{59}$$

or

$$\langle \delta c|\, [\mathbb{K}_{2H} - \epsilon\, \mathbb{M}_{2H}]\, |c\rangle = 0 \tag{60}$$

with real symmetric matrices

$$\mathbb{K}_{2H} = \int_0^R \{\rho^2 |B(\rho)\rangle\langle B(\rho)| - 2\rho |N(\rho)\rangle\langle N(\rho)|\} d\rho \tag{61}$$

$$\mathbb{M}_{2H} = \int_0^R \rho^2 |N(\rho)\rangle\langle N(\rho)| d\rho \tag{62}$$

Notice from Eq. (61) that $\mathbb{K}_{2H}$ is not positive definite. With arbitrary virtual spline coefficients, the generalized matrix eigenproblem becomes

$$[\mathbb{K}_{2H} - \epsilon\, \mathbb{M}_{2H}]\,|c\rangle = |0\rangle \tag{63}$$

Here, the eigen-energies $\epsilon$ are real and the eigenvectors of spline coefficients form an orthogonal set. However, the values of $\epsilon$ can become negative or zero, as expected.

Finally, we consider the fourth order formulation for the radial modes of the hydrogen atom, based on the weak form developed in Eq. (37). In discrete form, this can be written

$$\begin{aligned}\delta\mathcal{A}_{4H} \cong \int_0^R \{ &\rho^2 \,\langle\delta c|H(\rho)\rangle\langle H(\rho)|c\rangle + 2\rho\,\langle\delta c|H(\rho)\rangle\langle B(\rho)|c\rangle + 2\rho\,\langle\delta c|B(\rho)\rangle\langle H(\rho)|c\rangle \\ &+ 4(1-\rho)\,\langle\delta c|B(\rho)\rangle\langle B(\rho)|c\rangle + 4\,\langle\delta c|N(\rho)\rangle\langle N(\rho)|c\rangle \\ &- \epsilon^2\rho^2\,\langle\delta c|N(\rho)\rangle\langle N(\rho)|c\rangle\} \, d\rho - 2\,\langle\delta c|N(0)\rangle\langle N(0)|c\rangle = 0\end{aligned} \tag{64}$$

After operating with the transpose, bringing the spline coefficient vectors outside the integral, and identifying the system matrices, the following eigenproblem can be defined

$$[\mathbb{K}_{4H} - \epsilon^2\, \mathbb{M}_{4H}]\,|c\rangle = |0\rangle \tag{65}$$

where

$$\begin{aligned}\mathbb{K}_{4H} = \int_0^R \{\, &\rho^2\,|H(\rho)\rangle\langle H(\rho)| + 2\rho\,|H(\rho)\rangle\langle B(\rho)| + 2\rho\,|B(\rho)\rangle\langle H(\rho)| \\ &+ 4(1-\rho)\,|B(\rho)\rangle\langle B(\rho)| + 4\,|N(\rho)\rangle\langle N(\rho)|\} \, d\rho - 2\,|N(0)\rangle\langle N(0)|\end{aligned} \tag{66}$$

$$\mathbb{M}_{4H} = \int_0^R \rho^2\,|N(\rho)\rangle\langle N(\rho)|\, d\rho \tag{67}$$

Once again, the two system matrices in Eq. (65) and (66) are real and symmetric. However, Eq. (66) is in general indefinite.

Next, we present the RSM radial eigensolutions for the hydrogen atom. In terms of non-dimensional coordinates, a much more extensive domain is needed with a significantly greater number of spline intervals to capture accurate results, compared to the previous problem of the harmonic potential. In particular, the outer radial limit is set to $R = 250$ for both the 2nd and 4th-order analyses.

In the 2nd-order Schrodinger case, we take a total of $N = 500$ spline intervals using cubic ($k = 4$) splines with $C^0$ continuity. The resulting RSM eigenvalues are all real, having both negative and positive values, as expected from the analytical solution. The negative values correspond closely to the Rydberg energies associated with quantum numbers one through ten, as shown in Table 2, Notice that the RSM solutions are accurate to eight significant digits for the initial eight modes, but then begin to deteriorate for modes 9 and 10. Meanwhile, the positive $\epsilon$ values from the RSM analyses approximate the unbound continuous spectrum.

Solution of the 4th order Schrodinger equation RSM variational formulation is more challenging due to the singular nature of the potential $V(r)$ and the presence of the Dirac delta function in the weak form of Eq. (36). Results are presented below for $N = 1000$ intervals of $C^1$-continuous quartic splines ($k = 5$). The eigenvalues $\epsilon^2$ are all real, as expected. However, as explained previously, there is no guarantee that all eigenvalues will be positive.

Table 2. Computed radial eigenenergies of the hydrogen atom with the 2nd-order Schrodinger equation

| Quantum Number $Q_n$ | Rydberg energy $\epsilon$ (2nd Order Exact) | Rydberg energy $\epsilon$ (2nd Order RSM) | Effective Quantum Number $Q_n$ (2nd Order RSM) |
|---|---|---|---|
| 1 | -1 | -0.99999986 | 1.00000007 |
| 2 | -1/4 | -0.24999999 | 2.00000003 |
| 3 | -1/9 | -0.11111111 | 3.00000003 |
| 4 | -1/16 | -0.06250000 | 4.00000002 |
| 5 | -1/25 | -0.04000000 | 5.00000002 |
| 6 | -1/36 | -0.02777778 | 6.00000002 |
| 7 | -1/49 | -0.02040816 | 7.00000002 |
| 8 | -1/64 | -0.01562500 | 8.00000000 |
| 9 | -1/81 | -0.01234579 | 8.99995833 |
| 10 | -1/100 | -0.01001815 | 9.99093594 |

Numerical results show that there is in fact a single negative eigenvalue, say $\epsilon_0^2 < 0$, due to the finiteness of the computational domain $\rho = \in [0, R]$. The corresponding purely imaginary eigenenergy $\epsilon$ goes to zero, as $R \to \infty$. The remaining eigenvalues are positive with one portion approximating the continuous spectrum of unbound states having $\epsilon > 0$ and the other portion representing the discrete bound states associated with quantum numbers $Q_n = n$ for which $\epsilon < 0$. The difference in eigenenergy between adjacent modes $\epsilon_{m+1} - \epsilon_m$ can be used to separate the unbound and bound modes, as seen in Fig. 2. The appearance of each quantum number is clearly visible in that figure and can be used to identify the bound eigenstates. Table 3 provides the exact Rydberg energies for each of the first ten quantum numbers, along with the RSM computational results. Again, the numerical solutions are accurate to many significant digits with some loss of precision for $Q_9$ and $Q_{10}$.

The analytical form of the first three purely radial mode shapes for $Q_1$, $Q_2$ and $Q_3$, orthonormal with respect to $\rho$, are written:

$$\psi_{100}(\rho) = C_1 e^{-\rho} \tag{68a}$$

$$\psi_{200}(\rho) = C_2 (2 - \rho) e^{-\rho/2} \tag{68b}$$

$$\psi_{300}(\rho) = C_3 (27 - 18\rho + 2\rho^2) e^{-\rho/3} \tag{68c}$$

with $C_1 = 2$, $C_2 = \sqrt{2}/2$ and $C_3 = 2(\sqrt{3})^{-7}$.

Figure 3 displays both the 2$^{nd}$ and 4$^{th}$ order eigenfunctions obtained from the RSM analyses in comparison with the exact solutions from Eq. (68). Excellent agreement is achieved in all cases.

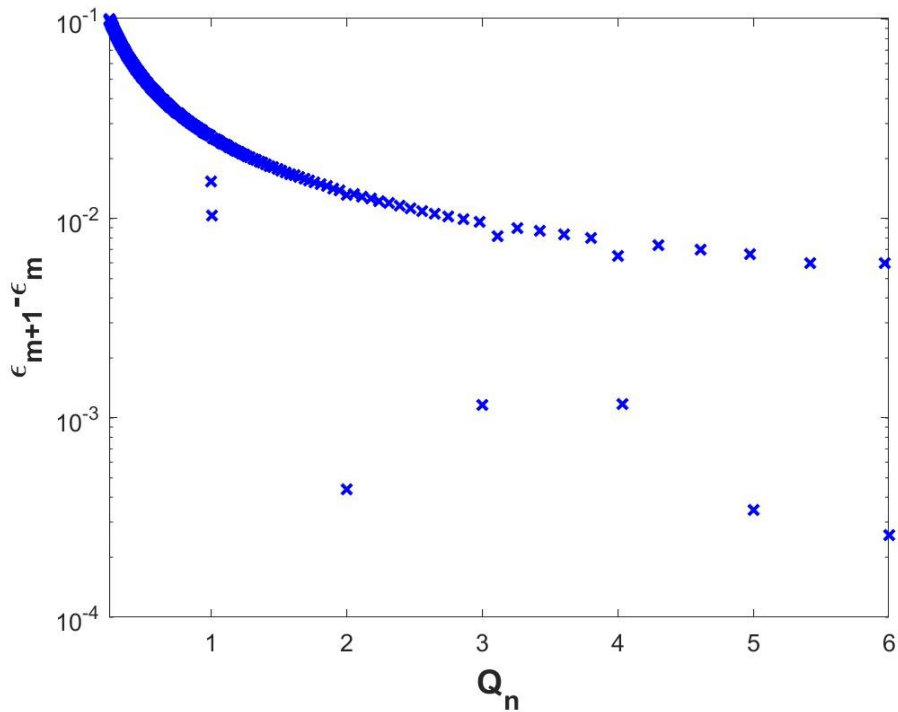

Figure 2. 4$^{th}$-order hydrogen atom radial eigenenergy separations (RSM results)

Table 3. Computed radial eigenenergies of the hydrogen atom with the 2$^{nd}$ and 4$^{th}$-order Schrodinger matter-wave equations

| Quantum Number | Rydberg energy $\epsilon$ (2$^{nd}$ Order Exact) | Rydberg energy $|\epsilon|$ (4$^{th}$ Order RSM) | Effective Quantum Number (4$^{th}$ Order RSM) |
|---|---|---|---|
| 1 | -1 | 1.00000000 | 1.00000000 |
| 2 | -1/4 | 0.25000000 | 2.00000000 |
| 3 | -1/9 | 0.11111111 | 3.00000000 |
| 4 | -1/16 | 0.06250000 | 4.00000002 |
| 5 | -1/25 | 0.04000000 | 5.00000002 |
| 6 | -1/36 | 0.02777778 | 5.99999996 |
| 7 | -1/49 | 0.02040816 | 6.99999989 |
| 8 | -1/64 | 0.01562500 | 7.99999990 |
| 9 | -1/81 | 0.01234593 | 8.99990822 |
| 10 | -1/100 | 0.01001632 | 9.99184982 |

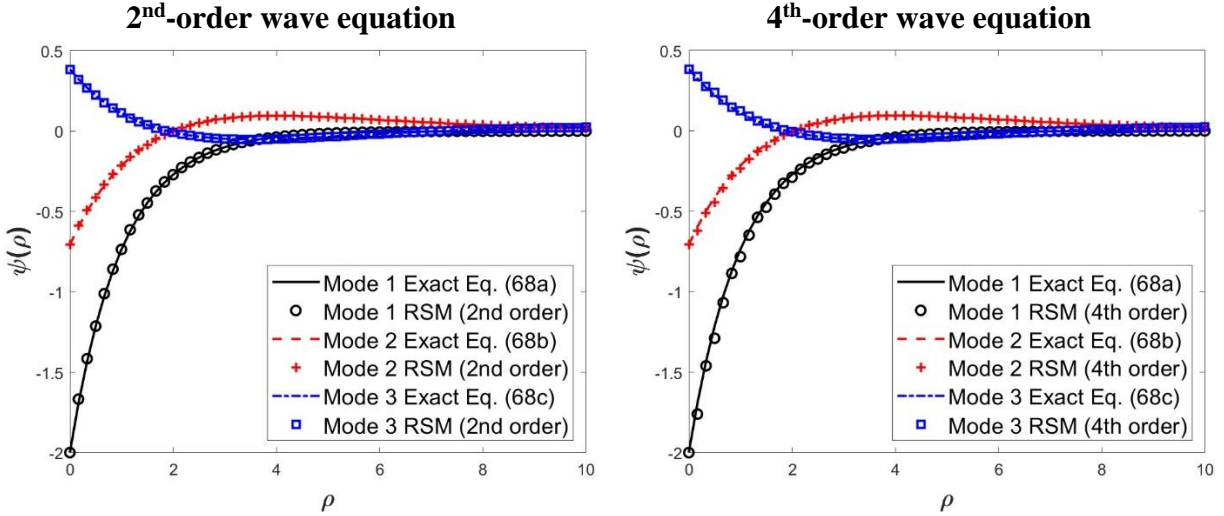

Figure 3. The first 3 eigenmodes of the Hydrogen atom. Exact 2$^{nd}$-order solutions (lines) versus the computational results with the Ritz Spline Method (points) implemented to the 2$^{nd}$ order and 4$^{th}$ order descriptions

**Conclusions**

In his seminal part IV, Annalen der Physik vol. 81, 1926 paper [1], Schrödinger has developed a clear understanding about the wave equation that produces the correct quadratic dispersion relation for matter-waves and he first presents a real-valued wave equation that is 4$^{th}$-order in space and 2$^{nd}$-order in time. Schrödinger proceeds by splitting the 4$^{th}$-order real operator into the product of two, second-order, conjugate complex operators and retains only one of the two complex operators to construct his iconic 2$^{nd}$-order in space and 1$^{st}$-order in time, complex-valued wave equation. Within the course of a few months upon the publication of his 1926 paper [1], it appears that Schrödinger's initial discomfort with the 2$^{nd}$-order in space and 1$^{st}$--order in time, complex-valued matter-wave equation was eventually removed and any attempt to develop a real-valued description by the pioneers of quantum mechanics was soon abandoned [4-9].

In this paper, using a variational formulation, we show that Schrödinger's 4$^{th}$-order, real-valued matter-wave equation which involves the spatial derivatives of the potential $V(r)$, produces the precise eigenvalues of Schrödinger's 2$^{nd}$-order, complex-valued matter-wave equation together with an equal number of negative, mirror eigenvalues. By taking the spatial derivatives of the potential $V(r)$ as dictated by the 4$^{th}$-order real operator (see equation (4)), all terms in the variational statements developed in this paper are symmetric respect to $\delta\psi(\rho)$ and $\psi(r)$ and involve at most second-order derivatives. Our paper proceeds by developing discrete representations of the variational statements which can be written exclusively in terms of real, symmetric matrices and lead to matrix eigenproblems with real eigenvalues and sets of corresponding real orthogonal eigenvectors as is done in classical structural dynamics.

Accordingly, our paper concludes that there is a real-valued description of non-relativistic quantum mechanics in association with the existence of negative (repelling) energy levels. Schrödinger's classical 2$^{nd}$-order, complex-valued matter-wave equation which was constructed upon factoring the 4th-order, real-valued differential operator and retaining only one of the two conjugate complex operators is a simpler description of the matter-wave, since it does not involve the spatial derivatives of the potential $V(r)$, at the expense of missing the negative (repelling) energy levels.